\documentstyle[aps,eqsecnum,preprint,tighten]{revtex}

\begin{document}
\draft

%%%%%%%%%%%%%%%%%%%%%%%%%%%%%%%%%%%%%%%%%%%%%%%%%%%%%%%%%%%%%%%%%%%%%%%%%%%%%%

\preprint{\vbox{Submitted to Phys.\ Rev.\ C \hfill DOE/ER/40762--032\\
                \null\hfill UMPP \#94--118}}

\title{QCD sum rules for $\Sigma$ hyperons in nuclear matter}
\author{Xuemin Jin and Marina Nielsen%
\thanks{Permanent address: Instituto de F\'{\i}sica,
Universidade de S\~ao Paulo, 01498 - SP- Brazil.}}
\address{Department of Physics and Center for Theoretical Physics\\
University of Maryland, College Park, Maryland 20742}
\date{\today}
\maketitle
\begin{abstract}
Within finite-density QCD sum-rule approach we investigate the
self-energies of $\Sigma$ hyperons propagating in nuclear matter from a
correlator of $\Sigma$  interpolating fields evaluated in the nuclear
matter ground state. We find that the Lorentz vector self-energy of the
$\Sigma$ is similar to the  nucleon vector self-energy.  The magnitude
of Lorentz scalar self-energy of the $\Sigma$ is also close to the
corresponding value for nucleon; however, this prediction is sensitive
to the strangeness content of the nucleon and to the assumed density
dependence of certain four-quark condensate.  The scalar and vector
self-energies tend to cancel, but not completely.  The implications for
the couplings of $\Sigma$ to the scalar and vector mesons in nuclear
matter and for the $\Sigma$ spin-orbit force in a finite nucleus are
discussed.
\end{abstract}

%%%%%%%%%%%%%%%%%%%%%%%%%%%%%%%%%%%%%%%%%%%%%%%%%%%%%%%%%%%%%%%%%%%%%%%%%%%%

\section{INTRODUCTION}
\label{intro}

Large and canceling isoscalar Lorentz scalar and vector self-energies
for propagating nucleons in nuclear matter (unlike the conventional
nonrelativistic picture) are essential for the successes of
relativistic nuclear phenomenology~\cite{wallace1,hama1,serot1,jong1}.
It has been shown recently, within a finite-density QCD sum-rule
approach, that this physics might be motivated from quantum
chromodynamics (QCD) \cite{cohen1,cohen3,furnstahl1,jin1,jin2}. (Other
applications of sum rule methods to finite density problems are
discussed in
Refs.~\cite{drukarev1,henley1,hatsuda1,adami1,hatsuda2,asakawa}).
The predictions of QCD sum rules for the nucleon self-energies are
found to be consistent with those obtained from relativistic
phenomenological models (e.g., the relativistic optical potentials of
Dirac phenomenology~\cite{wallace1,hama1} or Brueckner
calculations~\cite{serot1,jong1}). However, it was found in
Refs.~\cite{furnstahl1,jin2} that the sum-rule predictions for the
scalar self-energy are sensitive to assumptions made about the density
dependence of certain four-quark condensates.  Clearly, definite
conclusions are not yet justified and further tests of the approach are
needed.  To this end, one is naturally led to consider the hyperons in
nuclear matter within the same framework.

By studying the hyperons, one can use both  experimental data on
hypernuclei and relativistic phenomenology to confront with the QCD
sum-rule predictions. Various
investigators\cite{brockmann,noble1,yamazaki1,%
jcohen,jennings,chiapparini1,chiapparini2,jcohen2,mares,mares1,%
glendenning1,rufa,schaffner} have applied the relativistic
phenomenology to hypernuclear physics. In these relativistic models,
the hyperons are coupled to the same scalar and vector fields as the
nucleon, but with different coupling strengths.  (These couplings are
also of great relevance to other branch of
physics\cite{glendenning2}.)  However, these strengths are not
well-established. QCD sum-rule predictions for the scalar and vector
self-energies may provide valuable insight into these couplings in the
nuclear medium.

In Ref.~\cite{jin3}, the self-energies of a $\Lambda$ hyperon
propagating in nuclear matter have been studied using the
finite-density QCD sum-rule methods. The sum-rule calculations indicate
that the self-energies of the $\Lambda$ are only about $\case{1}{3}$ of
the corresponding nucleon self-energies, suggesting a significant
deviation from SU(3). In this paper, we further extend the
finite-density QCD sum-rule methods to explore the self-energies of
$\Sigma$ hyperons in an infinite nuclear matter.

One of the compelling successes of relativistic models in describing
nucleon-nucleus interactions is the naturally large spin-orbit force
for nucleons in a finite nucleus.  This force depends on the
derivatives of the scalar and vector optical potentials, which add
constructively.  An analogous prediction for the $\Lambda$ hyperon
would seem to be problematic, if one adopts the {\it naive} SU(3)
prediction that each coupling for the $\Lambda$ should be $\case{2}{3}$
the coupling for the
nucleon\cite{pirner,dover1,jennings,chiapparini1,chiapparini2,jcohen2,mares,mares1}.
 [That is, if one assumes that the scalar ($\sigma$) and vector
($\omega$) mesons couple exclusively to the up and down quarks and not
to the strange quark.] In the $\Lambda$-nucleus system, recent
experiments indicate that the spin-orbit force is small, and even
consistent with zero~\cite{bouyssy,millener}.  This has raised
questions about the validity of relativistic nuclear phenomenology for
the hyperons.

In Ref.~\cite{brockmann}, a weak $\Lambda$  spin-orbit force was
achieved by taking the potentials (i.e., the couplings) for the
$\Lambda$ to be a factor of three smaller than for the nucleon. More
recently it has been
suggested\cite{noble1,jennings,chiapparini1,chiapparini2,jcohen2,mares,mares1}
that larger values of the scalar and vector coupling strengths,
consistent with the naive SU(3) prediction, can be used if a new tensor
coupling between hyperons
 and the vector meson ($\omega$) is introduced.  In
Refs.~\cite{jennings,jcohen2,mares}, it was argued that a quark-model
picture implies that the tensor couplings of the hyperons
($\Lambda$,$\Sigma$,$\Xi$) to the vector meson differ in
their magnitudes and signs and hence their contribution to the
spin-orbit force is different.
For $\Lambda$, this picture leads to a tensor coupling with strength
equal in magnitude to the corresponding vector coupling, but with the
opposite sign.  The net result, in combination with the scalar
contribution, is a small spin-orbit force for the $\Lambda$. However,
the experimental information available is not sufficient to resolve the
effects of the tensor couplings.  The sum-rule calculations in
Ref.~\cite{jin3} suggest that the coupling of the $\Lambda$ to mesons
are only about $\case{1}{3}$ of the corresponding nucleon couplings,
implying a significant deviation from the naive SU(3) prediction and a
weak $\Lambda$ spin-orbit force; the tensor coupling, however, does not
appear in the sum-rule calculations in a uniform nuclear matter.

At the moment, experimental evidence of $\Sigma$ hypernuclei is
insufficient.  However, a number of
authors\cite{jcohen2,mares,mares1,glendenning1} have extended the
relativistic phenomenology to $\Sigma$ hypernuclei and presented
theoretical predictions.  In Refs.~\cite{jcohen2,mares,mares1}, the
naive SU(3) prediction for the coupling strengths of $\Sigma$ to the
scalar and vector mesons were adopted (the vector coupling for $\Sigma$
is $\case{2}{3}$ the coupling for nucleon and the scalar coupling for
$\Sigma$ is slightly smaller than $\case{2}{3}$ the coupling for
nucleon); in addition the tensor coupling between the $\Sigma$ and the
vector meson  was included. In the quark-model picture of
Refs.~\cite{jcohen2,dover1,jennings,mares}, the tensor coupling of
$\Sigma$ to $\omega$ has the {\it same} sign as the corresponding
vector coupling, in contrast to the $\Lambda$ case.
With the quark model values for the tensor coupling, the spin-orbit
force for the $\Sigma$ was found to be comparable with the nucleon
spin-orbit force. In Ref.~\cite{glendenning2} the tensor coupling was
omitted and universal couplings were assumed for {\it all} hyperons;
the ratio of $\Sigma$ to $\Lambda$ spin-orbit force is about $0.9$.
The sum-rule approach
may offer independent information on the scalar and vector couplings
for $\Sigma$ and new insight into deviations from SU(3) and into the
spin-orbit force, although the prediction of a tensor coupling is not
tested in the calculations described here.

The finite-density QCD sum-rule approach focuses on a correlation
function of interpolating fields, made up of quark fields, which  carry
the quantum numbers of the hadron of interest. The correlation function
is evaluated in the ground state of nuclear matter instead of the QCD
vacuum (as in the usual sum rules).  For spin-${1\over 2}$ baryons,
this function can,   in general, be decomposed into three invariant
functions of two kinematic invariants.  The appearance of an additional
invariant function compared to the vacuum case is due to an additional
four-vector at finite density, the four-velocity of the nuclear
medium.  In the rest frame of the medium, the analytic properties of
these invariant functions can be studied through a Lehmann
representation in energy.

The quasibaryon excitations (i.e., the quasiparticle excitations with
baryon quantum numbers) are characterized by the discontinuities of the
invariant functions across the real axis, which we use to identify the
on-shell self-energies.  By introducing a simple phenomenological
ansatz for these spectral densities, we obtain a representation of the
correlation function.  On the other hand, the correlation function can
be evaluated at large spacelike momenta using an operator product
expansion (OPE). By equating these two different representations using
appropriately weighted integrals, one obtains QCD sum rules, which
relate the baryon self-energies in the nuclear medium to QCD Lagrangian
parameters and finite-density condensates \cite{furnstahl1}.

We find that the finite-density QCD sum rules predict strong Lorentz
scalar and vector self-energies for the $\Sigma$ hyperon.  In
particular, the $\Sigma$ vector self-energy  is very similar to the
sum-rule prediction for the nucleon vector self-energy.  In terms of an
effective theory of baryons and mesons, this implies larger couplings
than would be indicated by SU(3) symmetry and the quark model.  The
predictions for the vector self-energy are essentially determined by
the nucleon density and the fact $\langle
s^{\dagger}s\rangle_{\rho_N}=0$, and the predicted ratio of the vector
self-energy to the zero-density $\Sigma$ mass is found to be largely
insensitive to the details of the calculation.  In contrast, the
predictions for the scalar self-energy are sensitive to the strangeness
content of the nucleon and to the assumed density dependence of
four-quark condensate $\langle\overline{q}q\rangle_{\rho_N}^{2}$.
If we assume that the nucleon has a significant strangeness content
{\it and} the four-quark condensate
$\langle\overline{q}q\rangle_{\rho_N}^{2}$ has weak density dependence,
then the magnitude of the $\Sigma$ scalar self-energy is found to be
close to the corresponding value for nucleon and tends to cancel the
vector self-energy, which is compatible with relativistic
phenomenology.  On the other hand,  if the strangeness content of the
nucleon is small or $\langle\overline{q}q\rangle_{\rho_N}^{2}$ depends
significantly on the nucleon density, the magnitude of $\Sigma$ scalar
self-energy is very small and the net $\Sigma$ self-energy is
sizable and repulsive.

This paper is organized as follows. The sum rules for  $\Sigma$
hyperons in nuclear matter are established in Sec.~\ref{sumrule}. The
results are presented in Sec.~\ref{results} and discussions are given
in Sec.~\ref{discussion}. Section \ref{summary} is a summary.

%%%%%%%%%%%%%%%%%%%%%%%%%%%%%%%%%%%%%%%%%%%%%%%%%%%%%%%%%%%%%%%%%%%%%%%
%
\section{Finite-density sum rules for $\Sigma$}
\label{sumrule}

In this section, we derive the finite-density sum rules for $\Sigma$.
We use the methods developed in Refs.~\cite{furnstahl1,jin1}, and refer
the reader to these papers for further details. We work to leading
order in perturbation theory. (The leading-logarithmic perturbative
corrections are included through anomalous dimension factors.) In the
OPE for $\Sigma$ correlator, we consider all condensates up to
dimension four, and the terms up to first order in the strange quark
mass $m_s$.  Contributions proportional to the up and down current
quark mass are neglected since they give numerically small
contributions. In addition, we include the contributions from the
dimension-six four-quark condensates (in the first order of $m_s$, we
only consider the terms which do not vanish in the zero-density
limit).  All other dimension six and higher-dimensional condensates are
neglected since we expect their contributions to be small in the region
of optimal Borel mass.

QCD sum rules for $\Sigma$ at finite density study the correlator
defined by
\begin{equation}
\Pi_\Sigma(q)\equiv i\int d^4 x\, e^{iq\cdot x}
\langle\Psi_0|{\rm T}[\eta_\Sigma(x)\overline{\eta}_\Sigma(0)]|\Psi_0\rangle\ ,
\label{corr_def}
\end{equation}
where $\eta_\Sigma(x)$ is a colorless interpolating field, constructed
from quark fields, which carries the quantum numbers of $\Sigma$. The
ground state of nuclear matter $|\Psi_0\rangle$ is characterized by the
nucleon density $\rho_N$ in the rest frame and the nuclear matter
four-velocity $u^{\mu}$; it is assumed to be invariant under parity and
time reversal. Here we consider the baryon interpolating fields that
contain no derivatives and couple to spin-${1\over 2}$ states only.
There are two linearly independent fields with these features. For the
nucleon, Ioffe~\cite{ioffe1,ioffe2} has argued that the optimal choice
is $\eta_{N}(x)$=$\epsilon_{abc}\left[u_a^{\mbox{\tiny
T}}(x)C\gamma_\mu u_b(x)\right] \gamma_5\gamma^\mu d_c(x)\ $ (for  the
proton).  The interpolating field for $\Sigma^+$ can be
directly obtained by an SU(3)-transformation of $\eta_{N}(x)$
\cite{reinders1}:
\begin{eqnarray}
\eta_{\Sigma}(x)&=&\epsilon_{abc}
\left[u_a^{\mbox{\tiny T}}(x)C\gamma_\mu u_b(x)\right]
\gamma_5\gamma^\mu s_c(x) ,
\label{intfield}
\end{eqnarray}
where T denotes a transpose in Dirac space, $C$ is the charge
conjugation matrix, $a$, $b$, and $c$ are color indices, and $u(x)$
 and $s(x)$ are the up and strange quark fields, respectively.  The
analogous interpolating field for $\Sigma^-$ follows by changing the up
quark field into down quark field. In the calculations to follow, we
assume the isospin symmetry and use $q$ to denote either up or down
quark field. The interpolating field Eq.~(\ref{intfield}) will be used
in our calculations.

Lorentz covariance, parity and time reversal then imply that the
correlator can only have three distinct structures
\cite{cohen1,furnstahl1,jin1}:
\begin{equation}
\Pi_\Sigma(q)\equiv\Pi_s(q^2,q\!\cdot\!u)+\Pi_q(q^2,q\!\cdot\!u)\rlap{/}{q}
+\Pi_u(q^2,q\!\cdot\!u)\rlap{/}{u}\ .
\end{equation}
[A potential invariant function multiplying $(q_\mu u_\nu-q_\nu
u_\mu)\sigma^{\mu\nu}$ vanishes due to time-reversal invariance and a
potential invariant function multiplying $ \epsilon^{\lambda\rho\mu\nu}
(q_\lambda u_\rho-q_\rho u_\lambda)\sigma_{\mu\nu}$ vanishes due to
parity conservation]. The three invariant functions, $\Pi_s$, $\Pi_q$
and $\Pi_u$, are functions of the two Lorentz scalars $q^2$ and
$q\!\cdot\!u$.
 In the zero-density limit, $\Pi_u\rightarrow 0$, and $\Pi_s$ and
$\Pi_q$ become functions of $q^2$ only.  For convenience, we will work
in the rest frame of nuclear matter hereafter, where $u^{\mu}=(1, {\bf
0})$ and $q\!\cdot\!u\rightarrow q_0$; we also take
$\Pi_i(q^2,q\!\cdot\!u)\rightarrow\Pi_i(q_0,|{\bf q}|)$
($i=\{s,q,u\}$). To obtain QCD sum rules, we need to construct a
phenomenological representation for $\Pi_\Sigma(q)$ and to evaluate
$\Pi_\Sigma(q)$ using  OPE techniques.

In the rest frame of nuclear matter, the analytic properties of
$\Pi_\Sigma(q)$ can be made manifest by a Lehmann
representation\cite{fetter}, which leads to a dispersion relation in
$q_0$ of the form \cite{furnstahl1}
\begin{equation}
\Pi_i(q_0,|{\bf q}|)={1\over 2\pi i}\int^{\infty}_{-\infty}
    d\omega{\Delta \Pi_i(\omega,|{\bf q}|)\over \omega-q_0}
          +\mbox{polynomial}
\label{des_re}
\end{equation}
for each invariant function $\Pi_i$, $i=\{s,q,u\}$. The polynomial
stands for contributions from the contour at large $|q_0|$, which will
be eliminated by a subsequent Borel transform (see below).  The
discontinuity $\Delta \Pi_i$ (which is the spectral density up to a
constant)  defined by
\begin{equation}
\Delta\Pi_i(\omega,|{\bf q}|)\equiv \lim_{\epsilon\rightarrow 0^+}
  [\Pi_i(\omega+i\epsilon,|{\bf q}|)-\Pi_i(\omega-i\epsilon,|{\bf q}|)]\,,
\label{discon_def}
\end{equation}
contains the spectral information on the quasiparticle, quasihole, and
higher energy states.

In QCD sum-rule applications, one parametrizes the spectral density
with a small number of spectral parameters characterizing the
resonances in the channel of interest (e.g., poles, residues, etc.). In
vacuum, the spectral weights for baryon and antibaryon are related by
charge conjugation symmetry and one usually parametrizes the spectral
density as a single sharp pole representing the lowest resonance plus a
smooth continuum representing higher mass states. At finite density,
the ground state is no longer invariant under ordinary charge
conjugation. Thus, the spectral densities for baryon and antibaryon are
not simply related.

The width of $\Sigma$ in free space is small and can be
ignored on hadronic scales. At finite density, the width of the
$\Sigma$ will be broadened due to strong conversions. Here we assume that
the broadened width is relatively small on the hadronic scale and that
a quasiparticle description of the $\Sigma$ is reasonable. In the
context of relativistic phenomenology, the $\Sigma$ is assumed to
couple to the same scalar and vector fields as the nucleons in the
nuclear matter, and are treated as  quasiparticles with real Lorentz
scalar and vector self-energies. We follow Ref.~\cite{furnstahl1}
and assume a pole approximation for the quasibaryon and take the
spectral ansatz to be (higher-energy states are included in a continuum
contribution, as discussed below)
\begin{eqnarray}
\Delta\Pi_s(\omega,|{\bf q}|)&=&-2\pi
i{M^\ast_\Sigma\lambda^{\ast^2}_\Sigma\over 2 E^\ast_q}
\left[\delta(\omega-E_q)
-\delta(\omega-\overline{E}_q)\right]\ ,
\label{ansz_s}
\\*
\Delta\Pi_q(\omega,|{\bf q}|)&=&-2\pi
i{\lambda^{\ast^2}_\Sigma\over 2E^\ast_q}\left[\delta(\omega-E_q)
-\delta(\omega-\overline{E}_q)\right]\ ,
\label{ansz_q}
\\*
\Delta\Pi_u(\omega,|{\bf q}|)&=&+2\pi
i{\Sigma_v\lambda^{\ast^2}_\Sigma\over 2E^\ast_q}
\left[\delta(\omega-E_q)
-\delta(\omega-\overline{E}_q)\right]\ ,
\label{ansz_u}
\end{eqnarray}
where $\lambda^{\ast^2}_\Sigma$ is an overall residue. Here we have
defined $M^\ast_\Sigma\equiv M_\Sigma+\Sigma_s$, $E^\ast_q\equiv
\sqrt{M^{\ast^2}_\Sigma+{\bf q}^2}$, $E_q\equiv \Sigma_v+
\sqrt{M^{\ast^2}_\Sigma+{\bf q}^2}$, and $\overline{E}_q\equiv
\Sigma_v-\sqrt{M^{\ast^2}_\Sigma+{\bf q}^2}$, where $M_\Sigma$ is the
mass of $\Sigma$ and $\Sigma_s$ and $\Sigma_v$ are identified as the
scalar and vector self-energies of a $\Sigma$ hyperon in  nuclear
matter. The positive- and negative-energy poles are at $E_q$ and
$\overline{E}_q$, respectively. Since we want to focus on the
positive-energy $\Sigma$ pole, we approximate the spectral functions
for the positive-energy $\Sigma$ by a quasiparticle pole while
suppressing contributions from the region of the negative-energy
excitations. This is achieved by manipulating the parts of the
correlator that are even and odd in $q_0$ (see below).

We now turn to the QCD expansion of $\Pi_\Sigma(q)$, which is obtained
by applying the operator product expansion to the time-ordered operator
product in Eq.~(\ref{corr_def}).  In the present formalism, the
correlator is studied in the limit that $q_0$ becomes large and
imaginary while $|{\bf q}|$ remains fixed (in the nuclear matter rest
frame).  This limit takes $q^2 \rightarrow -\infty$ with
$|q^2/q\!\cdot\!u|\rightarrow \infty$, which satisfies the conditions
discussed in Ref.~\cite{collins} for a short distance expansion.  At
finite density, the OPE for the invariant functions of
spin-$\case{1}{2}$ baryon correlator takes the general
form\cite{furnstahl1,jin1}
\begin{equation}
\Pi_i(q^2,q\!\cdot\!u)=\sum_n C^i_n(q^2,q\!\cdot\!u)
\langle\widehat O_n\rangle_{\rho_N}\ ,
\label{blib2}
\end{equation}
where $\langle{\widehat
O_n}\rangle_{\rho_N}\equiv\langle\Psi_0|{\widehat O_n}|\Psi_0\rangle$.
The $C^i_n(q^2,q\!\cdot\!u)$ ($i=\{{\rm s},q,u\}$) are the Wilson
coefficients, which depend on QCD Lagrangian parameters such as the
quark masses and the strong coupling constant.  The most important
feature is that the local composite operators are defined such that all
density dependence of the correlator resides in the condensates; the
Wilson coefficients are then independent of
density~\cite{jin1,furnstahl1}. The operators $\widehat O_n$ are
ordered by dimension (measured as a power of mass) and the
$C^i_n(q^2,q\!\cdot\!u)$ for higher-dimensional operators fall off by
corresponding powers of $Q^2\equiv -q^2$.  Therefore, for sufficiently
large $Q^2$, the operators of lowest dimension dominate, and the OPE
can be truncated after a small number of lower-dimensional operators.

The Wilson coefficients only depend on $q^\mu$, and the ground-state
expectation values of the operators are proportional to tensors
constructed from the nuclear matter four-velocity $u^\mu$, the metric
$g^{\mu\nu}$, and the antisymmetric tensor
$\epsilon^{\kappa\lambda\mu\nu}$.  In Eq.~(\ref{blib2}) we incorporate
the contraction of $q^\mu$ (from the OPE) and $u^\mu$ (from the
ground-state expectation values of the operators) into the definition
of the Wilson coefficients $C^i_n(q^2,q\!\cdot\!u)$.  Thus the
dependence on $q\!\cdot\!u$ is solely in the form of polynomial
factors.  We have also suppressed the dependence on the normalization
point $\mu$.

The Wilson coefficients can be calculated using the  fixed-point gauge
\cite{fock1,schwinger1}  and  standard background-field techniques
\cite{cronstrom1,smilga1,hubschmid1,novikov3,shifman2,reinders1}.  To
obtain the Wilson coefficients one can apply Wick's theorem to the
coordinate space time-ordered product in Eq.~(\ref{corr_def}),
retaining only those contributions in which the quark fields are fully
contracted, and using the quark propagators in the presence of the
nonperturbative nuclear medium, given in Refs.~\cite{jin1,jin3}, for
the contractions.

Although the four-quark condensates have dimension six, their
contribution to the baryon correlator are particularly important since
the corresponding Wilson coefficients do not carry the large numerical
suppression factors typically associated with loops~\cite{shifman1}.
By using the quark propagators given in Refs.~\cite{jin1,jin3} in the
calculations, one includes the contributions from the four-quark
condensates automatically with the condensates in their in-medium
factorized form~\cite{furnstahl1,jin1}. (See Sec.~\ref{results} for
discussion of the factorization approximation).

For convenience we split the invariant function $\Pi_i(q_0,|{\bf q}|)$
into two pieces that are even and odd in $q_0$:
\begin{equation}
\Pi_i(q_0,|{\bf q}|)=\Pi^{\mbox{\tiny\rm E}}_i(q_0^2,|{\bf q}|)+q_0
\Pi^{\mbox{\tiny\rm O}}_i(q_0^2,|{\bf q}|)\ ,
\\*
\label{inv_sep}
\end{equation}
for $i=\{s,q,u\}$. The results of our calculations are
\begin{eqnarray}
\Pi_s^{\mbox{\tiny\rm E}}
&=&-{m_s\over 32\pi^4}(q^2)^2 \ln(-q^2)+{1\over 4\pi^2 }q^2
\ln(-q^2)
\langle\overline{s}s\rangle_{\rho_N}
-{4m_s\over 3\pi^2}{q_{0}^2\over q^2}
\langle q^\dagger iD_0 q\rangle_{\rho_N}
\nonumber\\*
& &\null
-{4m_s \over 3q^2}\langle\overline{q}q\rangle_{\rho_N}^{2}\, ,
\label{PisE_ope_s}
\\*
%\end{eqnarray}
%
%\begin{eqnarray}
\Pi_s^{\mbox{\tiny\rm O}}
&=&{m_s\over 2\pi^2}\ln(-q^2)\left(\langle q^\dagger q\rangle_{\rho_N}
-\langle s^\dagger s\rangle_{\rho_N}\right)
-{4\over 3q^2}\langle\overline{s}s\rangle_{\rho_N}\langle
q^\dagger q\rangle_{\rho_N}  ,
\label{PisO_ope_s}
\\*
%\end{eqnarray}
%
%\begin{eqnarray}
\Pi_q^{\mbox{\tiny\rm E}}
&=&-{1\over 64\pi^2}(q^2)^{2}\ln(-q^2)
-{1\over 32\pi^2}\ln(-q^2)\left<{\alpha_s\over\pi}G^2\right>_{\rho_N}
\nonumber\\*[7.2pt]
& &\null
   -{1\over 144\pi^2}\left(\ln(-q^2)-4{q_0^2\over q^2}\right)
\left<{\alpha_s\over\pi}\left[(u\!\cdot\!
G)^2+(u\!\cdot\!\widetilde{G})^2\right]\right>_{\rho_N}
\nonumber\\*
& &\null
-{m_s\over 18\pi^2}\left(5\ln(-q^2)-2{q_0^2\over
q^2}\right)\langle\overline{s}s\rangle_{\rho_N}+{4\over 9\pi^2}
\left(\ln(-q^2)-{q_{0}^2\over q^2}\right)\langle q^\dagger
iD_0 q\rangle_{\rho_N}
\nonumber\\*[7.2pt]
& &\null
+{1\over 9\pi^2}\left(\ln(-q^2)-4{q_{0}^2\over q^2}\right)
\langle s^\dagger iD_0 s\rangle_{\rho_N}
-{2\over 3q^2}\langle\overline{q}q\rangle_{\rho_N}^{2}
\nonumber\\*
& &\null
-{4\over 3q^2}\langle q^\dagger q\rangle_{\rho_N}\langle s^\dagger
s\rangle_{\rho_N}\, ,
\label{PiqE_ope_s}
\\*
%\end{eqnarray}
%
%\begin{eqnarray}
\Pi_q^{\mbox{\tiny\rm O}}
&=&{1\over 6\pi^2}\ln(-q^2)\left(\langle q^\dagger q\rangle_{\rho_N}
+\langle s^\dagger s\rangle_{\rho_N}\right)\, ,
%\nonumber\\*
%& &\null
%-{8\over 27(q^2)^2}\langle q^\dagger
%q\rangle_{\rho_N}\left(4m_s\langle\overline{s}s\rangle_{\rho_N}-\langle
%s^\dagger iD_0 s\rangle_{\rho_N}\right)-{16m_s\over
%9(q^2)^2}\langle\overline{q}q\rangle_{\rho_N}\langle s^\dagger
%s\rangle_{\rho_N} \, ,
\label{PiqO_ope_s}
\\*[7.2pt]
%\end{eqnarray}
%
%\begin{eqnarray}
\Pi_u^{\mbox{\tiny\rm E}}
&=&
{1\over 12\pi^2}q^2\ln(-q^2)
\left(7\langle q^\dagger q\rangle_{\rho_N}
+\langle s^\dagger s\rangle_{\rho_N}\right)\, ,
%+{8m_s\over 9q^2}\langle\overline{q}q\rangle_{\rho_N}\langle s^\dagger
%s\rangle_{\rho_N}
%\nonumber\\*
%& &+{4m_s\over 27q^2}\left(1+10{q_{0}^2\over q^2}\right)\langle q^\dagger
%q\rangle_{\rho_N}\langle \overline{s}s\rangle_{\rho_N}
%\nonumber\\*
%& &
%+{4\over 27q^2}\left(17-40{q_{0}^2\over q^2}\right)\langle q^\dagger
%q\rangle_{\rho_N}\langle s^\dagger iD_0 s\rangle_{\rho_N}\, ,
\label{PiuE_ope_s}
%\\*
\end{eqnarray}
\begin{eqnarray}
\Pi_u^{\mbox{\tiny\rm O}}
&=&{1\over 9\pi^2}\ln(-q^2)\left(m_s\langle\overline{s}s\rangle_{\rho_N}
-16\langle q^\dagger iD_0 q\rangle_{\rho_N}-4\langle
s^\dagger iD_0 s\rangle_{\rho_N}\right)
\nonumber\\*
& &\null
+{1\over 36\pi^2}\ln(-q^2)\left<{\alpha_s\over\pi}
\left[(u\!\cdot\!     G)^2
+(u\!\cdot\!\widetilde{G})^2\right]\right>_{\rho_N}
\nonumber\\*
& &\null
-{4\over 3q^2}\left(\langle q^\dagger q\rangle_{\rho_N}^2
+\langle q^\dagger q\rangle_{\rho_N}\langle
s^\dagger s\rangle_{\rho_N}\right)\, ,
\label{PiuO_ope_s}
\end{eqnarray}
where we have introduced the notation $G^2\equiv G^{\mbox{\tiny\rm
{A}}}_{\mu\nu}G^{{\mbox{\tiny\rm {A}}}\mu\nu}$ and
$\widetilde{G}^{{\mbox{\tiny\rm  {A}}}\kappa\lambda} \equiv{1\over
2}\epsilon^{\kappa\lambda\mu\nu}G^{\mbox{\tiny\rm {A}}}_{\mu\nu}$, and
have omitted the polynomials in $q^2$ and $q_0^2$, which
vanish under the Borel transform.
%
%Due to the isospin symmetry
Here we take
$\langle\overline{q}\widehat{O}q\rangle_{\rho_N}\equiv(\langle\overline{
u}\widehat{O}u\rangle_{\rho_N}+
\langle\overline{d}\widehat{O}d\rangle_{\rho_N})/2$, with
$\widehat{O}$ a combination of Dirac matrices, gluon field tensors, and
covariant derivatives.
In Eqs.~(\ref{PisE_ope_s}--\ref{PiuO_ope_s}),
we have neglected the terms proportional to $\langle\overline{q}iD_0
q\rangle_{\rho_N}$ since $\langle\overline{q}iD_0 q\rangle_{\rho_N}
={3\over 4}(m_u+m_d)\rho_N\simeq 0$ (see Ref.~\cite{jin1}).

In the present calculations, we have neglected all the terms
proportional to the dimension-five quark and quark-gluon condensates.
There is no difficulty in calculating the Wilson coefficients for these
condensates. However, we have no reliable approach to determine these
condensates. With the estimated values of the quark and quark-gluon
condensates (containing only light quarks) given in Ref. \cite{jin1},
the dimension-five quark and quark-gluon condensates make numerically
small contributions to the nucleon sum rules~\cite{jin2}.  We expect
that the contributions from the dimension-five quark and quark-gluon
condensates are also small in the $\Sigma$  sum rules, so we neglect
these terms.  Contributions from three-gluon condensates, which are
dimension six, are also expected to be numerically small compared to
those of four-quark condensates since the four-quark condensates enter
at tree level whereas the three-gluon condensates involve loops
\cite{jin1}.

The QCD sum rules follow by equating the spectral representation of the
correlator to the corresponding OPE representation. To improve the
overlap of the two descriptions for QCD sum rules for hadrons in
vacuum, one typically applies a Borel transform to both sides of the
sum rules.  For  practical purposes in the present approach, the Borel
transform can be applied using the operator ${\cal B}$ defined by
\begin{eqnarray}
{\cal B}[f(Q^2,|{\bf q}|)]&\equiv&\lim_{Q^2,n\rightarrow\infty}
{(Q^2)^{n+1}\over n!}\left(-{\partial\over \partial Q^2}\right)^n
f(Q^2,|{\bf q}|)\equiv \hat{f}(M^2,|{\bf q}|)\ ,
\nonumber\\*
Q^2&\equiv&-q_{0}^{2},\hspace{2cm} M^2\equiv{Q^2\over n}.
\label{borel_def}
\end{eqnarray}
The result $\hat{f}(M^2,|{\bf q}|)$ depends on the Borel mass $M$.

The finite-density QCD sum rules are then given
by~\cite{furnstahl1,jin2}
\begin{equation}
{\cal B}[\Pi^{\mbox{\tiny\rm E}}_i(q_0^2,|{\bf q}|)-\overline{E}_q
\Pi^{\mbox{\tiny\rm O}}_i(q_0^2,|{\bf q}|)]_{\rm QCD}
 ={\cal B}[\Pi^{\mbox{\tiny\rm E}}_i(q_0^2,|{\bf q}|)-\overline{E}_q
\Pi^{\mbox{\tiny\rm O}}_i(q_0^2,|{\bf q}|)]_{\rm phen}\ ,
\\*
\label{sum_def}
\end{equation}
for $i=\{s,q,u\}$, where the left-hand side is obtained from the OPE
and right-hand side from the phenomenological dispersion relations. To
see how this manipulation suppresses  contributions from the region of
negative-energy excitations, we rewrite Eq.~(\ref{des_re}) as
\begin{equation}
\Pi_i(q_0,|{\bf q}|)={1\over 2\pi i}\int_{-\infty}^{\infty}d\omega\,
\omega\, {\Delta\Pi_i(\omega,|{\bf q}|)\over \omega^2-q_0^2}
+{q_0\over 2\pi i}\int_{-\infty}^{\infty}d\omega \,
{\Delta\Pi_i(\omega,|{\bf q}|)\over \omega^2-q_0^2}\,,
\\*
\label{des_rere}
\end{equation}
where we have omitted the polynomial term. We then obtain
\begin{equation}
{\cal B}[\Pi^{\mbox{\tiny\rm E}}_i(q_0^2,|{\bf q}|)-\overline{E}_q
\Pi^{\mbox{\tiny\rm O}}_i(q_0^2,|{\bf q}|)]_{\rm phen}
={1\over 2\pi i}\int_{-\infty}^{\infty}d\omega \,
(\omega-\overline{E}_q)\Delta\Pi_i(\omega,|{\bf q}|)\,
e^{-\omega^2/M^2} \,.\
\\*
\label{des_mani}
\end{equation}
The negative-energy pole contribution is now suppressed by the factor
$(\omega-\overline{E}_q)$, which equals  zero at
$\omega=\overline{E}_q$. For Borel $M$ near the $\Sigma$ energy,
contributions of higher-energy states to the integral are exponentially
suppressed.

Perturbative corrections $\sim\alpha_s^n$ can be taken into account in
the leading logarithmic approximation through anomalous-dimension
factor, $L\equiv {\ln{M/\Lambda_{\rm QCD}}\over\ln{\mu/\Lambda_{\rm
QCD}}}$~\cite{ioffe1},
where
%
%After the Borel transform, the effect of these corrections is
%to multiply each term appearing on the OPE side of the sum rules by the
%factor~\cite{ioffe1}
%
%\begin{equation}
%(L)^{-2\Gamma_{\eta} +\Gamma_{ O_{n}}}
%\equiv\left({\ln{M/\Lambda_{\rm QCD}}\over\ln{\mu/\Lambda_{\rm QCD}}}
%\right)^{-2\Gamma_{\eta}+\Gamma_{O_{n}}}\ ,
%\end{equation}
%
%
%where $\Gamma_{\eta}$ is the anomalous dimension of the interpolating
%field $\eta_{\Sigma}$, $\Gamma_{O_{n}}$ is the anomalous dimension of
%the corresponding local operator (including the current quark masses),
$\mu$ is the normalization point of the operator product expansion and
$\Lambda_{\rm QCD}$ is the QCD scale parameter. In our numerical
calculations, we take $\mu=0.5\,\text{GeV}$ and $\Lambda_{{\rm
QCD}}=0.1\,\text{GeV} \cite{ioffe3}$.
%
%The anomalous dimensions
%$\Gamma_{\eta}$ and $\Gamma_{O_{n}}$ are dependent on the number of
%flavor, $N_f$ \cite{ioffe1,peshkin}; we take $N_f=3$.

Finally, the contributions from higher-energy states are roughly
approximated  using the leading terms in the OPE, starting at an
effective threshold; these contributions can be included by modifying
the terms with positive powers of $M^2$ on the OPE side of each
sum-rule equation as follows\cite{ioffe1,furnstahl1,jin2}:
\begin{eqnarray}
M^2&\rightarrow&M^2E_0\equiv M^2\left(1-e^{-s_0^*/M^2}\right)\ ,
\label{con_0}
\\*
M^4&\rightarrow&M^2E_1\equiv M^4\left[1-e^{-s_0^*/M^2}
\left({s_0^*\over M^2}+1\right)\right]\ ,
\label{con_1}
\\*
M^6&\rightarrow&M^6E_2\equiv M^6\left[1-e^{-s_0^*/M^2}\left({s_0^{*^2}\over
2M^4}
+{s_0^{*^2}\over M^2}+1\right)\right]\ ,
\label{con_2}
\end{eqnarray}
where we define the continuum threshold $s_0^*=\omega_0^2-{\bf q}^2$,
with $\omega_0$ the energy at the continuum threshold. In principle,
the effective thresholds are different for positive and negative
energies and for the different sum rules.  The former differences are
critical in some sum rule formulations \cite{asymmetry}, but are not
numerically important in the present formulation.  Furthermore, the
thresholds are relatively poorly determined by the sum rules and
effects due to different thresholds in different sum rules may be
absorbed  by slight changes in the other parameters.  In the present
paper, we use a universal effective threshold for simplicity.

With the spectral ansatz Eqs.~(\ref{ansz_s})--(\ref{ansz_u}) and the
OPE results of Eqs.~(\ref{PisE_ope_s})--(\ref{PiuO_ope_s}), we obtain
the finite-density sum rules for the $\Sigma$:
\begin{eqnarray}
\lambda^{\ast^2}_\Sigma M^{\ast}_\Sigma e^{-(E_q^2-{\bf q}^2)/M^2}
&=&{m_s\over 16\pi^4}M^{6}E_2 L^{-8/9}
-{M^4\over 4\pi^2}E_1\langle\overline{s}s\rangle_{\rho_N}
\nonumber\\*
& &\null
+{m_s\over 2\pi^2}\overline{E}_qM^2E_0
\left(\langle q^{\dagger}q\rangle_{\rho_N}
-\langle s^{\dagger}s\rangle_{\rho_N}\right)L^{-8/9}
\nonumber\\*
& &\null
+{4m_s\over 3\pi^2}{\bf q}^2\langle q^{\dagger}
iD_0 q\rangle_{\rho_N}
L^{-8/9}
+{4m_s\over 3\pi^2}\langle\overline{q}q\rangle^{2}_{\rho_N}
\nonumber\\*
& &\null
-{4\over 3}\overline{E}_q\langle\overline{s}s\rangle_{\rho_N}
\langle q^{\dagger}q\rangle_{\rho_N}\,,
\label{sum_s}
%\end{eqnarray}
\\*
%\begin{eqnarray}
%
\lambda^{\ast^2}_\Sigma e^{-(E_q^2-{\bf q}^2)/M^2}
&=&{M^6\over 32\pi^4}E_2L^{-4/9}+{M^2\over 32\pi^2}
\left<{\alpha_s\over\pi}G^2\right>_{\rho_N}E_0L^{-4/9}
\nonumber\\*
& &\null
+{M^2\over 144\pi^2}\left(E_0-4{{\bf q}^2\over M^2}\right)
\left<{\alpha_s\over\pi}\left[(u\!\cdot\!
G)^2+(u\!\cdot\!\widetilde{G})^2\right]\right>_{\rho_N}L^{-4/9}
\nonumber\\*
& &\null
+{m_s\over 18\pi^2}M^2\left(5E_0-2{{\bf q}^2\over M^2}\right)
\langle\overline{s}s\rangle_{\rho_N}L^{-4/9}
\nonumber\\*
& &\null
-{4M^2\over 9\pi^2}\left(E_0-{{\bf q}^2\over M^2}\right)
\langle q^{\dagger} iD_0 q\rangle_{\rho_N}L^{-4/9}
\nonumber\\*
& &\null
-{M^2\over 9\pi^2}\left(E_0-4{{\bf q}^2\over M^2}\right)
\langle s^{\dagger} iD_0 s\rangle_{\rho_N}L^{-4/9}
\nonumber\\*
& &\null
+{\overline{E}_q\over 6\pi^2}M^{2}E_0
\left(\langle q^{\dagger} q\rangle_{\rho_N}
+\langle s^{\dagger} s\rangle_{\rho_N}\right)L^{-4/9}
\nonumber\\*
& &\null
+{2\over 3}
\langle\overline{q}q\rangle_{\rho_N}^2
L^{4/9}
+ {4\over 3}
\langle q^\dagger q\rangle_{\rho_N}
\langle s^\dagger s\rangle_{\rho_N}L^{-4/9}\, ,
\label{sum_q}
\end{eqnarray}
%\\*
\begin{eqnarray}
\lambda^{\ast^2}_\Sigma\Sigma_v e^{-(E_q^2-{\bf q}^2)/M^2}
&=&{1\over 12\pi^2}M^4E_1\left(7\langle q^{\dagger}
q\rangle_{\rho_N}+
\langle s^{\dagger} s\rangle_{\rho_N}\right)L^{-4/9}
%+{8m_s\over 9}\langle\overline{q}q\rangle_{\rho_N}\langle s^{\dagger}
%s\rangle_{\rho_N}L^{-4/9}
\nonumber\\*
& &\null
-{\overline{E}_q\over 9\pi^2}M^{2}E_0\left(m_s\langle\overline{s}s
\rangle_{\rho_N}
-16\langle q^{\dagger} iD_0 q\rangle_{\rho_N}
-4\langle s^{\dagger} iD_0 s\rangle_{\rho_N}\right)L^{-4/9}
\nonumber\\*
& &\null
-{\overline{E}_q\over 36\pi^2}M^{2}E_0\left<{\alpha_s\over\pi}
\left[(u\!\cdot\!
G)^2+(u\!\cdot\!\widetilde{G})^2\right]\right>_{\rho_N}L^{-4/9}
\nonumber\\*
& &\null
+{4\overline{E}_q\over 3}\langle q^{\dagger} q\rangle_{\rho_N}
\left(\langle q^{\dagger} q\rangle_{\rho_N}
+\langle s^\dagger s\rangle_{\rho_N}\right)L^{-4/9}\ .
\label{sum_u}
\end{eqnarray}
Here $E_0$, $E_1$ and $E_2$  have been defined in
Eqs.~(\ref{con_0})--(\ref{con_2}). We have ignored the anomalous
dimensions of dimension four operators, either because the operators
are renormalization-group invariant, because the anomalous dimension is
small, because the corresponding condensates give small contributions,
or because the accuracy to which the nucleon matrix elements of the
operators are known is such that anomalous-dimension corrections
represent an unwarrantable refinement (see Sec.~\ref{results}).  We
will adopt the values of the corresponding condensates at the scale of
$1\,\text{GeV}$ in our numerical calculations.  The four-quark
operators are in general not renormalization-covariant, so they mix
with one another under the renormalization group~\cite{shifman1}.  In
vacuum, the anomalous dimension effects do not violate the
factorization assumption to within $10\%$ \cite{shifman1}, and thus one
usually assumes that the anomalous dimension of a four-quark operator
is equal to the sum of the anomalous dimensions of the factorized
operators \cite{ioffe1,belyaev1}. In the present paper, we follow this
assumption.

%%%%%%%%%%%%%%%%%%%%%%%%%%%%%%%%%%%%%%%%%%%%%%%%%%%%%%%%%%%%%%%%%%%%%%%%%%

%%%%%%%%%%%%%%%%%%%%%%%%%%%%%%%%%%%%%%%%%%%%%%%%%%%%%%%%%%%%%%%%%%%%%%%%%%%%%
%
\section{Results}
\label{results}

%%%%%%%%%%%%%%%%%%%%%%%%%%%%%%%%%%%%%%%%%%%%%%%%%%%%%%%%%%%%%%%%%%%%%%%%
%

\subsection{In-medium condensates}
\label{result_cond}

To obtain the predictions for the $\Sigma$ self-energies from the
finite-density sum rules derived in the previous section, we need to
know the in-medium condensates appearing in the sum rules. To first
order in the nucleon density, one can write
\begin{equation}
\langle\hat{O}\rangle_{\rho_N}=\langle\hat{O}\rangle_{\rm vac}+
\langle\hat{O}\rangle_N\rho_N +\cdot\cdot\cdot \, ,
\end{equation}
where $\cdot\cdot\cdot$ denotes correction terms that are of higher
order in $\rho_N$, and $\langle\hat{O}\rangle_N$ is the spin averaged
nucleon matrix element.  (Note that this is not a Taylor series in
$\rho_N$.) For a general operator $\widehat{O}$ there is not a
systematic way to study contributions to
$\langle\widehat{O}\rangle_{\rho_N}$ that are of higher order in
$\rho_N$.  Model-dependent estimates in Ref.~\cite{cohen3} suggest that
the linear approximation to $\langle\overline{q}q\rangle_{\rho_N}$
should be  good (higher-order corrections $\sim 20\%$ of the linear
term) up to nuclear matter saturation density.  Here we assume the
first-order approximation of {\it all\/} condensates to be reasonable
for calculating scalar and vector self-energies up to nuclear matter
saturation density. Justifying the limits of this type of density
expansion is an important topic for further study.

The simplest in-medium condensates are $\langle
q^{\dagger}q\rangle_{\rho_N}$ and $\langle
s^{\dagger}s\rangle_{\rho_N}$. Since the baryon current is conserved,
$\langle q^{\dagger}q\rangle_{\rho_N}$ is proportional to the (net)
nucleon and strangeness densities:  $\langle
q^{\dagger}q\rangle_{\rho_N}={3\over 2}\rho_N$, and $\langle
s^{\dagger}s\rangle_{\rho_N}=0$.  These are exact results.  The other
dimension three and four quark and gluon
condensates have been studied in
Refs.~\cite{cohen3,drukarev1,jin1,jin3}. The results are:
\begin{eqnarray}
& &\langle\overline{q}q\rangle_{\rho_N}=\langle\overline{q}q\rangle_{\rm vac}
+\langle\overline{q}q\rangle_N \rho_N=
\langle\overline{q}q\rangle_{\rm vac}+{\sigma_N\over 2m_q}\rho_N\ ,
\label{qbar_im}
\\*
& &\langle\overline{s}s\rangle_{\rho_N}=\langle\overline{s}s
\rangle_{\rm vac}
+\langle\overline{s}s\rangle_N\rho_N=\langle\overline{s}s
\rangle_{\rm vac}
+y{\sigma_N\over 2m_q}\rho_N\ ,
\label{sbar_im}
\\*
& &\left<{\alpha_s\over\pi}G^2\right>_{\rho_N}=
\left<{\alpha_s\over\pi}G^2\right>_{\rm vac}-(650\,\text{MeV})\rho_N\ ,
\\*
& &\left<{\alpha_s\over\pi}\left[(u^\prime\!\cdot\!
G)^2+(u^\prime\!\cdot\!\widetilde{G})^2\right]\right>_{\rho_N}
=-(100\,\text{MeV})\rho_N\ ,
\\*
& &\langle q^\dagger iD_0 q\rangle_{\rho_N}=(180\,\text{MeV})\rho_N\ ,
\\*
& &\langle s^{\dagger} iD_0
s\rangle_{\rho_N}=\case{m_s}{4}\langle\overline{s}s\rangle_{\rho_N}+
(18\,\text{MeV})\rho_N\ .
\label{partn-5}
\end{eqnarray}
where $m_{q}\equiv{1\over 2}(m_u+m_d)$ is the average of the up and
down current quark masses, $\sigma_N$ is the nucleon $\sigma$ term, and
$y\equiv
\langle\overline{s}s\rangle_N/\langle\overline{q}q\rangle_N$ is a real
parameter, measuring the strangeness content of the nucleon.

Four-quark condensates are numerically important in both the vacuum and
the finite-density baryon sum rules because they contribute in tree
diagrams and do not carry the numerical suppression factors typically
associated with loops.  In the sum rules derived in Sec.~\ref{sumrule},
we included the contributions from the four-quark condensates in their
in-medium factorized forms; however, the factorization approximation
may not be justified in nuclear
matter~\cite{furnstahl1,jin1,jin2,jin3}.
 In the nucleon sum rules, the ``scalar-scalar'' four-quark condensate
$\langle  \overline{q} q\rangle_{\rho_N}^2$ gives important
contributions; the sum-rule predictions are sensitive to the value of
this condensate \cite{furnstahl1,jin2}.  This four-quark condensate
also appears in the $\Sigma$ sum rules.  In its factorized form, this
scalar-scalar four-quark condensate has a strong density dependence;
one might suspect that this strong density dependence is an artifact of
the factorization approximation.  Thus we follow Ref.~\cite{jin2} and
parametrize the scalar-scalar four-quark condensate so that it
interpolates between its factorized form in free space and its
factorized form in nuclear matter:
\begin{eqnarray}
\langle\overline{q}q\rangle_{\rho_N}^2\longrightarrow
\langle\widetilde{\overline{q}q}\rangle_{\rho_N}^2&=&(1-f)
\langle\overline{q}q\rangle_{\rm vac}^2+f
\langle\overline{q}q\rangle_{\rho_N}^2\ ,
\label{4-quark_para}
\end{eqnarray}
where $f$ is a real parameter. The density dependence of the
scalar-scalar four-quark condensate is thus parametrized by $f$, and
the density dependence of $\langle\overline{q}q\rangle_{\rho_N}$  [see
Eqs.~(\ref{qbar_im})]. The factorized four-quark condensate
$\langle\overline{q}q\rangle_{\rho_N}^2$ appearing in
Eqs.~(\ref{sum_s}--\ref{sum_u}) will be replaced by
$\langle\widetilde{\overline{q}q}\rangle_{\rho_N}^2$ in the
calculations to follow.  Studies of nucleon self-energies with QCD sum
rules yield results in strong contradiction to experiment unless $0\leq
f\leq 0.5$~\cite{furnstahl1,jin2}; however, there is not an independent
determination of the density dependence.  Here we will consider only
this range of $f$.  The other four-quark condensates have much smaller
numerical contributions, so we just use the in-medium factorized form
for simplicity.

The values of $\sigma_N$ and $y$ remain
controversial~\cite{banerjee,gensini,gasser}. In this paper, we take
$\sigma_N=45\,\text{MeV}$, which is the value obtained in a recent
analysis~\cite{gasser}, and consider values of $y$ in the range of
$0-0.6$, which covers the values discussed
in~\cite{banerjee,gensini,gasser}. For the condensates in vacuum, we
use $\langle\overline{q}q\rangle_{\rm vac}\simeq -(245\,\text{MeV})^3$
($m_q\simeq 5.5\,\text{MeV}$) \cite{furnstahl1,jin2}, and take
$\langle\overline{s}s\rangle_{\rm
vac}=0.8\langle\overline{q}q\rangle_{\rm vac}$
\cite{belyaev1,reinders1,leinweber1} and
$\left<(\alpha_s/\pi)G^2\right>_{\rm vac}=(330\,\text{MeV})^4$
\cite{shifman1,hatsuda2}. We consider the strange quark mass $m_s$ in
the range of $100-200\,\text{MeV}$.

%%%%%%%%%%%%%%%%%%%%%%%%%%%%%%%%%%%%%%%%%%%%%%%%%%%%%%%%%%%%%%%%%%%%%%%%%%%%
%

\subsection{Sum-rule analysis}
\label{result_anay}

In principle, the predictions based on the sum rules should be
independent of the auxiliary parameter $M^2$.  In practice, however, we
have to truncate the OPE and use a simple phenomenological ansatz for
the spectral density; thus one expects the two descriptions to overlap
only in some limited range of $M^2$ (at best).  As a result, one
expects to see a ``plateau'' in the predicted quantities as functions
of $M^2$.  The studies of the vacuum sum rules  for the octet baryons
and the nucleon sum rules at finite-density show that the sum rules
truncated at dimension-six condensates do not provide a particularly
convincing plateau~\cite{ioffe1,leinweber1,furnstahl1}.\footnote%
{Including direct-instanton effects in nucleon sum rules in vacuum
leads to a more convincing plateau \cite{dorokhov,forkel}.}
Nevertheless, we will assume that the sum rule actually has a region of
overlap, although imperfect.  We follow
Refs.~\cite{furnstahl1,jin2,jin3} and rely on the cancellation of
systematic discrepancies by normalizing all finite-density
self-energies to the zero-density prediction for the mass.  One hopes
that this might compensate for general limitations of the sum rules.
All the finite-density results presented are obtained at nuclear matter
saturation density, which is taken to be $\rho_N=(110\,\text{MeV})^3$.

To analyze the sum rules and extract the self-energies, we sample the
sum rules in the fiducial region of $M^2$, where the contributions from
the highest-order condensates included in the sum rule are small and
the continuum contribution is controllable. Here we choose $1.0\leq
M^2\leq 1.6\,\text{GeV}^2$ as the optimization region [the ratios of
the self-energies to the $\Sigma$ mass are insensitive to the choice of
the upper bound of the optimal region (see Fig.~\ref{fig-7})]. The
study of the $\Sigma$ sum rules in vacuum suggests that the sum rules are
valid in this region~\cite{belyaev1,leinweber1}. To quantify the fit of
the left- and right-hand sides, we use the logarithmic
measure~\cite{belyaev1,leinweber1,furnstahl1,jin2,jin3}
\begin{equation}
\delta(M^2)=\ln\left[{\mbox{maximum}}\{\lambda^{\ast^2}_\Sigma
e^{-(E_q^2-{\bf q}^2)/M^2},
\Pi_s^{\prime}/M^{\ast}_\Sigma,
\Pi_q^{\prime},
\Pi_u^{\prime}/\Sigma_v\}\over{\mbox{minimum}}
\{\lambda^{\ast^2}_\Sigma e^{-(E_q^2-{\bf q}^2)/M^2},
\Pi_s^{\prime}/M^{\ast}_\Sigma,
\Pi_q^{\prime},\Pi_u^{\prime}/\Sigma_v\}\right]\ ,
\end{equation}
which is averaged over  $150$ points evenly spaced within the fiducial
region of $M^2$. Here $\Pi_{s}^{\prime}$, $\Pi_{q}^{\prime}$, and
$\Pi_{u}^{\prime}$ denote the right-hand sides of
Eqs.~(\ref{sum_s})--(\ref{sum_u}), respectively.
The predictions for $M^{\ast}_\Sigma$,
$\Sigma_v$, $\omega_0^2$, and $\lambda^{\ast^2}_\Sigma$ are obtained by
minimizing the averaged measure $\delta$. To get a prediction for the
$\Sigma$ mass in vacuum, we apply the same procedure to the sum rules
evaluated in the zero-density limit.

In Fig.~\ref{fig-1}, we displayed the optimized results for the ratios
$M^{\ast}_\Sigma/M_\Sigma$ and $\Sigma_v/M_\Sigma$ as functions of $y$
for $m_s=150\,\text{MeV}$,  $|{\bf q}|=270\,\text{MeV}$, and three
different values of $f$.  (The momentum dependence of the self-energies
for momenta below the Fermi surface is very weak, see
Fig.~\ref{fig-6}). It can be seen that $\Sigma_v/M_\Sigma$ is
insensitive to both $y$ and $f$. However, the
$M^{\ast}_\Sigma/M_\Sigma$ varies rapidly with $y$ and $f$. Therefore,
the sum rule prediction for the scalar self-energy is {\it strongly}
dependent on the strangeness content of the nucleon and on the density
dependence of the four-quark condensate.  For $f=0$ and the values of
$y$ in the range $0.4\leq y\leq 0.6$, the predictions are:
\begin{eqnarray}
M^{\ast}_\Sigma/M_\Sigma &\simeq& \text{0.78--0.85}\, ,
\\*
\Sigma_v/M_\Sigma &\simeq& \text{0.18--0.19}\, .
\label{typ-results}
\end{eqnarray}
On the other hand, for $f=0$ and small values of $y$ ($0\leq y\leq
0.2$), we find $\Sigma_v/M_\Sigma\sim 0.18$ and
$M^{\ast}_\Sigma/M_\Sigma\simeq \text{0.92--0.98}$. As $f$ increases,
$M^{\ast}_\Sigma/M_\Sigma$ increases, which implies an even smaller
magnitude of the scalar self-energy. The predictions for the ratios
$\lambda^{\ast 2}_\Sigma/\lambda^2_\Sigma$ and $s_0^\ast/s_0$ also
depend on $y$ and $f$. For $f=0$ and large values of $y$
($0.4\leq y\leq 0.6$), both the continuum threshold and the residue
$\lambda^{\ast^2}_\Sigma$ are close to their corresponding vacuum
values. For $f=0$ and small values of $y$ $(0\leq y\leq 0.2)$, the
continuum threshold increases by about $20\%$ relative to the vacuum
value and the residue $\lambda^{\ast^2}_\Sigma$ increases by about
$50\%$ relative to its vacuum value. As $f$ increases, both the
continuum threshold and the residue increase.

One can see, from the sum rules Eqs.~(\ref{sum_s})--(\ref{sum_u}),  that
the ratios $\Pi_s^\prime/\Pi_q^\prime$ and $\Pi_u^\prime/\Pi_q^\prime$
give $M^\ast_\Sigma$ and $\Sigma_v$ as functions of Borel $M^2$, and
$\Pi_s^\prime/\Pi_q^\prime$ in the zero-density limit yields $M_\Sigma$
as a function of $M^2$. In Fig.~\ref{fig-2}, the ratios
$M^\ast_\Sigma/M_\Sigma$ and $\Sigma_v/M_\Sigma$ are plotted as
functions of $M^2$ for $y=0.5$ and various values of $f$, with $E_q$,
$\overline{E}_q$ and the continuum threshold fixed at their optimized
values. The curves for the ratios $M^\ast_\Sigma/M_\Sigma$ and
$\Sigma_v/M_\Sigma$ are flat, and thus imply a weak dependence of the
predicted ratios on $M^2$ (though the individual sum-rule predictions
before taking ratios are not flat).

We plot $\lambda^{\ast^2}_\Sigma e^{-(E_q^2-{\bf q}^2)/M^2}$,
$\Pi_s^{\prime}(M^2)/M^{\ast}_\Sigma$, $\Pi_q^{\prime}(M^2)$, and
$\Pi_u^{\prime}(M^2)/\Sigma_v$ as functions of $M^2$ for $y=0.5$ and
$f=0$  in Fig.~\ref{fig-3}(a), with the predicted values for
$M^{\ast}_\Sigma$, $\Sigma_v$, $\omega_0^2$, and
$\lambda^{\ast^2}_\Sigma$.  If the sum rules work well, one should
expect to see that the four curves coincide with each other. It is seen
that their $M^2$ dependence in the Borel region of interest turns out
to be equal up to 15\%.  The overlap of the corresponding vacuum sum
rules (i.e., the zero-density limit) is illustrated in
Fig.~\ref{fig-3}(b).  We observe that the quality of the overlap for
the finite-density sum rules is similar to that of the corresponding
sum rules in vacuum.

The sensitivity of our predictions to $m_s$ is illustrated in
Fig.~\ref{fig-4}, where $y$ and $|{\bf q}|$ are fixed at $0.5$ and
$270\,\text{MeV}$, respectively. The predictions for
$M^{\ast}_\Sigma/M_\Sigma$ and $\Sigma_v/M_\Sigma$ are largely
insensitive to changes in $m_s$, with a variation of
$\Sigma_v/M_\Sigma$ less than $10\%$ in the range of
$m_s=0.1-0.2\,\text{GeV}$.  Fig.~\ref{fig-6} shows the three-momentum
$|{\bf q}|$ dependence of the predicted ratios. The results are only
weakly dependent on three momentum in the range of $|{\bf
q}|=0-500\,\text{MeV}$. Finally, the dependence of the results on the
choice of the upper bound of the Borel window is shown in
Fig.~\ref{fig-7}. The two dashed curves are obtained using a fixed
Borel window at $1.0\,\text{GeV}^2\leq M^2\leq 1.6\,\text{GeV}^2$. The
two solid curves are the results obtained by taking an upper bound of
the  Borel window such that the continuum contributions to the
phenomenological sides do not exceed 50\% of the total phenomenological
contributions to the sum rules (i.e., the sum of the quasiparticle pole
and the continuum contributions) while  fixing the lower bound at
$1.0\,\text{GeV}^2$. We have used the same procedure in extracting the
$\Sigma$ mass in vacuum. We see that the two set of curves are almost
indistinguishable, implying that changing the upper limit of the
optimum Borel region does not affect the sum-rule predictions for the
two ratios.

%%%%%%%%%%%%%%%%%%%%%%%%%%%%%%%%%%%%%%%%%%%%%%%%%%%%%%%%%%%%%%%%%%%%%%%%
\section{Discussion}
\label{discussion}

We note that the sum-rule predictions for the scalar self-energy are
quite sensitive to the strangeness content of the nucleon and to the
undetermined density dependence of certain four-quark condensate.
Therefore, we cannot draw definite conclusions about the $\Sigma$
scalar self-energy at this point. Nevertheless, we emphasize that the
sum-rule prediction for the normalized  vector self-energy,
$\Sigma_v/M_\Sigma$, is apparently insensitive to the details of
calculations.  For typical values of the relevant condensates and other
input parameters, $\Sigma_v/M_\Sigma\sim 0.18-0.21$.  The
finite-density {\it nucleon\/} sum rules predict $\Sigma_v/M_N\sim
0.25-0.30$~\cite{furnstahl1,jin2}. Thus, we find
$(\Sigma_v)_\Sigma/(\Sigma_v)_N\sim 0.8-1.1$.

This result, if interpreted in terms of a relativistic hadronic model,
would imply that the coupling of the $\Sigma$ to the Lorentz vector
field is very similar to the corresponding nucleon coupling in the same
ratio.  This compares to the naive SU(3) prediction of $\case{2}{3}$,
which is obtained by assuming that the mesons couple directly to
constituent quarks~\cite{pirner,jennings,chiapparini1,chiapparini2},
and thus suggests a significant deviation from SU(3) symmetry in
nuclear matter.  This deviation can be attributed to two sources.
First, the nuclear matter ground state is not SU(3) symmetric due to
the absence of  net strangeness.  This leads to further deviation of
various condensates from SU(3) relative to the deviation in vacuum. The
second source originates from the baryon interpolating fields. In the
present work, we used the interpolating field defined in
Eq.~(\ref{intfield}), corresponding to an axial vector diquark,
composed of two up quarks,
coupled to a strange quark. However, these quarks contribute to the sum
rules differently. This can be seen from the leading-order terms in the
sum rules.  In Eq.~(\ref{sum_u}) there is an extra factor of seven
multiplying  $\langle q^{\dagger} q\rangle_{\rho_N}$ relative to the
term proportional to $\langle s^{\dagger} s\rangle_{\rho_N}$.   Since
the strange quark does not couple to the nuclear vector current (i.e.,
$\langle s^{\dagger} s\rangle_{\rho_N}=0$),  the leading-order term of
Eq.~(\ref{sum_u}), which sets the scale for the $\Sigma$ vector
self-energy, is very close to the corresponding leading term for the
nucleon.

The dependence of the $\Sigma$ scalar self-energy on $y$ comes mainly
from the leading-order term (proportional to
$\langle\overline{s}s\rangle_{\rho_N}$) of Eq.~(\ref{sum_s}) and the
parametrization Eq.~(\ref{sbar_im}).  If we assume that the nucleon has
a large strangeness content (i.e., $y\sim 0.4-0.6$) and the four-quark
condensate $\langle\overline{q}q\rangle_{\rho_N}^2$ depends only weakly
on the nucleon density (i.e, if $f\sim 0$), we find
$M^{\ast}_\Sigma/M_\Sigma\sim 0.77-0.84$, which implies
$\Sigma_s/M_\Sigma\sim -(0.16-0.23)$.  With the nucleon sum-rule
prediction $M_N^{\ast}/M_N\sim 0.65-0.70$~\cite{furnstahl1,jin2}, we
obtain $(\Sigma_s)_\Sigma/(\Sigma_s)_N\sim 0.6-1.0$.  In a hadronic
model, this implies again a coupling of the $\Sigma$ to the Lorentz
scalar field close to that for the nucleon. In this case, there is a
significant degree of cancellation between the scalar and vector
self-energies, which is compatible with that implemented in the
relativistic phenomenological models.

In contrast, if the strangeness content of the nucleon is small (i.e.,
$y\leq 0.2$) or  if $\langle\overline{q}q\rangle_{\rho_N}^2$ has a
significant dependence on the nucleon density, the predicted ratio
$M^{\ast}_\Sigma/M_\Sigma$ is close to unity, implying that the scalar
self-energy is very small. The predicted vector self-energy, on the
other hand, is still essentially the same as the nucleon vector
self-energy.  Thus, in this case the sum rules predict incomplete
cancellation, and hence a sizable repulsive net self-energy for the
$\Sigma$. This result contradicts with that of the relativistic
models.

We now turn to the $\Sigma$ spin-orbit force in a finite nucleus. In
the present approach, all sum-rule predictions are obtained for uniform
infinite nuclear matter; thus one cannot obtain a direct prediction for
the $\Sigma$ spin-orbit force.  However, we can still get some partial
information on the $\Sigma$ spin-orbit force by adopting an approach
of  Dirac phenomenology~\cite{serot1,chiapparini2}. In this approach,
the coordinate-space potentials entering the Dirac equation for a
baryon scattering from a finite nucleus are assumed to follow a Fermi
distribution with two overall potential depths (scalar and vector),
which are independent of nuclei, and which can be associated with the
self-energies in infinite nuclear matter.  The spin-orbit force follows
by recasting the Dirac equation in Schr\"{o}dinger form. The resulting
spin-orbit potential is proportional to the sum of the magnitudes of
the potential depths (that is, the scalar and vector self-energies)
multiplied by the derivative of the assumed Fermi
distribution~\cite{serot1}.

The sum-rule predictions for the $\Sigma$ scalar and vector
self-energies in the present paper imply that the $\Sigma$ spin-orbit
force in a nucleus is somewhat weaker than (but comparable with) that
felt by a nucleon, but much stronger than that felt by a $\Lambda$.
This is consistent with that obtained in
Refs.~\cite{jcohen2,mares,mares1}.  However, we emphasize that in
Refs.~\cite{jcohen2,mares,mares1} the scalar and vector couplings,
consistent with SU(3), have been adopted and it is the extra tensor
coupling of $\Sigma$ to the vector meson that enhances the spin-orbit
force. Our sum-rule calculations, on the other hand, suggest that it is
the strong scalar and vector couplings, deviated from the naive SU(3)
prediction, that lead to large spin-orbit force. We also note that our
sum-rule predictions in this paper and those in Ref.~\cite{jin3} for
$\Lambda$ do not agree with the universal coupling assumption (i.e.,
all hyperons couple to the scalar and vector fields with the same
strength) suggested in Ref.~\cite{glendenning1}.

%%%%%%%%%%%%%%%%%%%%%%%%%%%%%%%%%%%%%%%%%%%%%%%%%%%%%%%%%%%%%%%%%%%%%%%%%%%
%

%
\section{Summary}
\label{summary}

In this paper, we have applied finite-density QCD sum-rule methods to
investigate the self-energies of a $\Sigma$ hyperon in nuclear matter.
The approach focuses on a correlator of  $\Sigma$ interpolating fields,
evaluated in the nuclear matter ground state.  A QCD expansion of the
correlator is obtained by applying the operator product expansion.  We
retained the contributions from all condensates up through dimension
four and to first-order in the strange quark mass $m_s$, and we also
included the contributions from the four-quark condensates.  The
expansion requires new condensates not present at zero density, as well
as information  on the density dependence of condensates.
In the rest frame of the nuclear matter, a Lehmann representation for
the correlator, with fixed three-momentum, leads to a dispersion
relation in the energy variable.  A simple quasiparticle pole ansatz,
with real Lorentz scalar and vector self-energies, is assumed for the
spectral functions associated with the three invariant functions
comprising the $\Sigma$ correlator.  Contributions from higher-energy
states are roughly approximated by a perturbative evaluation of the
correlator, starting at an effective threshold.

The sum-rule analysis indicates that the $\Sigma$ vector self-energy is
similar to the corresponding nucleon self-energy.  If we interpret this
result in terms of a relativistic hadronic model, it implies that the
vector coupling for the $\Sigma$ is similar to the corresponding
nucleon coupling to the vector meson.  The vector self-energy
is largely insensitive to the details of calculations, and is
essentially determined by the nuclear matter density and the fact that
$\langle s^{\dagger} s\rangle_{\rho_N}=0$.

The sum-rule predictions for the $\Sigma$ vector self-energy, along
with that for the $\Lambda$ in Ref.~\cite{jin3}, indicate a significant
deviation from SU(3) in nuclear matter. This deviation is mainly due to
the violation of SU(3) in the baryon interpolating fields and due to
the violation of SU(3) in the nuclear matter ground state. At this
stage, it is unclear whether the violation of SU(3) in the baryon
interpolating fields is connected to the violation of SU(3) in the
baryon wave functions. Further study of the dependence of the deviation
on the choice of the interpolating field will be important in
understanding the deviation from SU(3).

The scalar self-energy is found to be quite sensitive to the
strangeness content of the nucleon and the assumed density dependence
of the four-quark condensate $\langle\overline{q}q\rangle_{\rho_N}^2$.
We parametrized the density dependence of this four-quark condensate
in terms of its factorized form in free space and its factorized form
in nuclear matter.  If the nucleon has a significant strangeness
content and if the four-quark condensate
$\langle\overline{q}q\rangle_{\rho_N}^2$ has a weak density dependence,
the sum rules predict strong scalar and vector self-energies. This is
qualitatively compatible with relativistic models.  On the other hand,
if the strangeness content of the nucleon is small or
$\langle\overline{q}q\rangle_{\rho_N}^2$ depends significantly on the
nucleon density, the sum rules predict a very small scalar self-energy
and a strong vector self-energy, which differs from the relativistic
models.  Clearly, further study of the strangeness content of the
nucleon and the four-quark condensates in the nuclear medium is very
important, along with analyses of the higher-order density dependence
of other condensates and the contributions from the condensates with
higher dimension.

The sum-rule predictions for the $\Sigma$ scalar and vector
self-energies seem to imply a somewhat weaker spin-orbit force for
$\Sigma$ in a nucleus than that for a nucleon, but much stronger than
that for a $\Lambda$. This is compatible with those obtained from
relativistic phenomenological models with an extra tensor coupling
between the hyperons and the vector meson. This possible tensor
contribution, however, cannot be observed in uniform nuclear matter.

It is a straightforward exercise to study the self-energies of $\Xi$
within the same framework. However, in the $\Xi$ sum rules, the leading
order term of $\Pi_u$ is small relative to the higher order terms.
Thus, it might be difficult to extract useful information since one does
not have much control over the values of the higher order terms.

\vspace{0.5in}

One of the authors (X.J.) would like to thank Byron Jennings for
pointing out this problem to him and for helpful discussions. X.J.
thanks J. Mares for useful conversations.   We thank R. J. Furnstahl
and T. D. Cohen for reading
the manuscript carefully and H. Forkel for useful comments.
X.J.\ acknowledges support from the US National Science Foundations
under grant No. PHY-9058487 and from the Department of Energy under
Grant No.\ DE-FG02-93ER-40762.  M.N. acknowledges the warm hospitality
and congenial atmosphere provided by the Nuclear Theory Group of the
University of Maryland and support from FAPESP-Brazil.

%%%%%%%%%%%%%%%%%%%%%%%%%%%%%%%%%%%%%%%%%%%%%%%%%%%%%%%%%%%%%%%%%%%%

%%%%%%%%%%%%%%%%%%%%%%%%%%%%%%%%%%%%%%%%%%%%%%%%%%%%%%%%%%%%%%%%%%%%%%%
%

%%%%%%%%%%%%%%%%%%%%%%%%%%%%%%%%%%%%%%%%%%%%%%%%%%%%%%%%%%%%%%%%%%%%%
%
\begin{figure}
\caption{Optimized sum-rule predictions for
$M^{\ast}_\Sigma/M_\Sigma$ and $\Sigma_v/M_\Sigma$ as functions of
$y$. The three curves correspond to $f=0$ (solid), $f=0.25$
(dashed), and $f=0.5$ (dotted). The other input parameters are
described in the text.}
\label{fig-1}
\end{figure}
\begin{figure}
\caption{Ratios $M^{\ast}_\Sigma/M_\Sigma$ and $\Sigma_v/M_\Sigma$
as functions of Borel $M^2$ for $y=0.5$, with optimized predictions
for $E_q$, $\overline{E}_q$, and the continuum thresholds. The three
curves correspond to $f=0$ (solid), $f=0.25$ (long-dashed),
and $f=0.5$ (dotted). The other input parameters are the same as in
Fig.~\protect{\ref{fig-1}}.}
\label{fig-2}
\end{figure}
\begin{figure}
\caption{(a) The left- and right-hand sides of the sum rules as
functions of Borel $M^2$ for $y=0.5$ and $f=0$, with the
optimized values for $M^{\ast}_\Sigma$, $\Sigma_v$, $s_0^\ast$, and
$\lambda^{\ast^2}_\Sigma$. The other parameters are the same as in
Fig.~\protect{\ref{fig-1}}.  The four curves correspond to
$\Pi_s^{\prime}/M^{\ast}_\Sigma$ (solid), $\Pi_q^{\prime}$ (dashed),
$\Pi_u^{\prime}/\Sigma_v$ (dot-dashed), and $\lambda^{\ast^2}_\Sigma
e^{-(E_q^2-{\bf q}^2)/M^2}$ (dotted).  (b) The left- and right-hand
sides of the corresponding vacuum sum rules. The three curves
correspond to $\Pi_s^{\prime}/M_\Sigma$ (solid), $\Pi_q^{\prime}$
(dashed), and $\lambda^{2}_\Sigma e^{-M_\Sigma^2/M^2}$ (dot-dashed)
at the zero-density limit, with the optimized values for $M_\Sigma$,
$s_0$, and $\lambda_\Sigma^2$.}
\label{fig-3}
\end{figure}
\begin{figure}
\caption{Optimized predictions for $M^{\ast}_\Sigma/M_\Sigma$ and
$\Sigma_v/M_\Sigma$ as functions of $m_s$ for $y=0.5$. The three
curves correspond to $f=0$ (solid), $f=0.25$ (dashed), and $f=0.5$
(dotted). The other input parameters are the same as in
Fig.~\protect{\ref{fig-1}}.}
\label{fig-4}
\end{figure}
%
%\begin{figure}
%\caption{Optimized predictions for $M^{\ast}_\Sigma/M_\Sigma$ and
%$\Sigma_v/M_\Sigma$ as functions of $y$ for $f_1=0.25$. The three
%curves correspond to $f_2=0$ (solid), $f_2=0.5$ (dashed), and $f_2=1.0$
%(dotted). The other input parameters are the same as in
%Fig.~\protect{\ref{fig-1}}.}
%\label{fig-5}
%\end{figure}
%
\begin{figure}
\caption{Three-momentum dependence of the predicted
$M^{\ast}_\Sigma/M_\Sigma$ and $\Sigma_v/M_\Sigma$ for $y=0.5$.
The three curves correspond to $f=0$ (solid), $f=0.25$ (dashed), and
$f=0.5$ (dotted). The other input parameters are the same as in
Fig.~\protect{\ref{fig-1}}.}
\label{fig-6}
\end{figure}
\begin{figure}
\caption{Optimized sum-rule predictions for
$M^{\ast}_\Sigma/M_\Sigma$ and $\Sigma_v/M_\Sigma$ as functions of
$f$, with $y=0.5$.  The other input parameters are the same as in
Fig.~\protect{\ref{fig-1}}. The solid curves correspond to the results
obtained by requiring the continuum contributions to be less than 50\%
in the fiducial Borel region and the dashed curves correspond to the
results obtained using a fixed Borel window at
$1.0\,\text{GeV}^2\leq M^2\leq 1.6\,\text{GeV}^2$.}
\label{fig-7}
\end{figure}

\begin{references}
%
\bibitem{wallace1}S.~J. Wallace,
Annu.\ Rev.\ Nucl.\ Part.\ Sci.\ {\bf 37}, 267 (1987), and references therein.
%%
\bibitem{hama1}S. Hama, B.~C. Clark, E.~D. Cooper, H.~S. Sherif,
and R.~L. Mercer,
Phys.\ Rev.\ C {\bf 41}, 2737 (1990).
%
\bibitem{serot1}B.~D. Serot and J.~D. Walecka, Adv.\ Nucl.\ Phys.\
{\bf 16},1(1986); B.~D. Serot, Indiana Univ.
preprint IU-NTC $\#$92-6(1992).
%
\bibitem{jong1}F.~de Jong and R. Malfliet, Phys.\ Rev.\  C {\bf 44},
998(1991).
%
%
\bibitem{cohen1}T.~D. Cohen, R.~J. Furnstahl, and D.~K. Griegel,
Phys.\ Rev.\ Lett. {\bf 67}, 961 (1991).
%
\bibitem{cohen3}T.~D. Cohen, R.~J. Furnstahl, and D.~K. Griegel,
Phys.\ Rev.\ C {\bf 45}, 1881 (1992).
%
\bibitem{furnstahl1}R.~J. Furnstahl, D.~K. Griegel, and T.~D. Cohen,
Phys.\ Rev.\ C {\bf 46}, 1507(1992).
%
\bibitem{jin1}X. Jin, T.~D. Cohen, R.~J. Furnstahl,
and  D.~K. Griegel,  Phys.\ Rev.\ C {\bf 47}, 2882(1993).
%
\bibitem{jin2}X. Jin, M.~Nielsen, T.~D. Cohen, R.~J. Furnstahl,
and  D.~K. Griegel, Phys.\ Rev.\ C {\bf 49}, 464(1994).
%
\bibitem{drukarev1}E.~G. Drukarev and E.~M. Levin,
Pis'ma Zh.\ Eksp.\ Teor.\ Fiz.\ {\bf 48}, 307 (1988)
[JETP Lett.\ {\bf 48}, 338 (1988)];
Zh.\ Eksp.\ Teor.\ Fiz.\ {\bf 95}, 1178 (1989)
[Sov.\ Phys.\ JETP {\bf 68}, 680 (1989)];
Nucl.\ Phys.\ {\bf A511}, 679 (1990);
{\bf A516}, 715(E) (1990);
Prog.\ Part.\ Nucl.\ Phys.\ {\bf 27}, 77 (1991).
%
\bibitem{henley1}E.~M. Henley, J. Pasupathy,
Nucl.\ Phys.\ {\bf A556}, 467 (1993).
%
\bibitem{hatsuda1}T. Hatsuda, H. H{\o}gaasen, and M. Prakash,
Phys.\ Rev.\ Lett.\ {\bf 66}, 2851 (1991).
%%
\bibitem{adami1}C. Adami and G.~E. Brown,
Z.\ Phys.\ A {\bf 340}, 93 (1991).
%
%
\bibitem{hatsuda2}T. Hatsuda and S.~H. Lee,
Phys.\ Rev.\ C {\bf 46}, R34 (1992).
%
\bibitem{asakawa}M. Asakawa and C.~M. Ko,
Nucl.\ Phys.\ {\bf A560}, 399 (1993);
Phys.\ Rev.\ C {\bf 48}, R526 (1993).
%
%\bibitem{wilson1}K.~G. Wilson,
%Phys.\ Rev.\ {\bf 179}, 1499 (1969).
%
%
\bibitem{brockmann}R. Brockmann and W. Weise, Phys.\ Lett.\ {\bf B69},
167 (1977).
%; J. Boguta and S. Bohrmann, Phys.\ Lett.\ {\bf 102B}, 93 (1981).
%
\bibitem{noble1}J.~V. Noble, Phys.\ Lett. {\bf B89}, 325 (1980).
%
\bibitem{yamazaki1}T. Yamazaki,
Phys.\ Lett.\ {\bf 160B}, 227 (1985);
Nucl.\ Phys.\ {\bf A446}, 467c (1985).
%
\bibitem{jcohen}J. Cohen and R.~J. Furnstahl,
Phys.\ Rev.\ C {\bf 35}, 2231 (1987).
%
\bibitem{jennings}B.~K. Jennings, Phys.\ Lett.\ {\bf B246}, 325(1990).
%
\bibitem{chiapparini1}A.~O. Gattone, M. Chiapparini, and E.~D. Izquierdo,
Phys.\ Rev.\ C {\bf 44}, 548 (1991)
%
\bibitem{chiapparini2}M. Chiapparini, A.~O. Gattone, and B.~K. Jennings,
Nucl.\ Phys.\ {\bf A529}, 589 (1991).
%
%
\bibitem{jcohen2}J. Cohen and H.~J.\ Weber, Phys.\ Rev.\ C {\bf 44},
            1181 (1991).
%
\bibitem{mares}J. Mares, Particles and Nuclei XIII International Conference,
Book of Abstracts Vol. 2, 661 (1993).
%
\bibitem{mares1}J. Mares and B. K. Jennings, TRIUMF preprint.
%
\bibitem{glendenning1}N. K. Glendenning {\it et al.}, Phys.\ Rev. C {\bf 48},
889 (1993).
%
\bibitem{rufa}M.~Rufa, J.~Schaffner, J.~Maruhn, H.~St\"ocker,
    W.~Greiner, and P.-G.~Reinhard, Phys.\ Rev.\ C {\bf 42},
        2469 (1990).
%
\bibitem{schaffner}J. Schaffner, C. Greiner, and H. St\"ocker, Phys.\
Rev.\ C {\bf 46}, 322(1992).
%
%
\bibitem{glendenning2}N. K. Glendenning, Astrophys. J., {\bf 293}, 470 (1985);
O. V. Maxwell, Astrophys. J., {\bf 316}, 691 (1987);
M. Prakash, J. M. Latimer, and G. J. Pethick, Astrophys. J., {\bf 390},
L77 (1992).
%
%
\bibitem{jin3}X. Jin and R.~J. Furnstahl,  Phys.\ Rev.\ C {\bf 49}, 1190(1994).
%
\bibitem{pirner}H.~J. Pirner, Phys.\ Lett. {\bf 85B}, 190(1979).
%
\bibitem{dover1}C. B. Dover and A. Gal, Prog.\ Part.\ Nucl.\ Phys. {\bf 12},
171(1984).
%
\bibitem{bouyssy}A. Bouyssy, Nucl.\ Phys.\ {\bf A290}, 324 (1977).
%
\bibitem{millener}D.~J. Millener, C.~B. Dover, and A. Gal, Phys.\ Rev.\ C
{\bf 38}, 2700 (1988); D.~J.\ Millener, A.~Gal, C.~B.\ Dover,
and R.~H. Dalitz,
Phys.\ Rev.\ C {\bf 31}, 499 (1985).
%
%\bibitem{exspin}R. Chrien, Nucl.\ Phys.\ {\bf A478}, 705c (1988);
%C. Milner {\it et al}, Phys.\ Rev.\ Lett. {\bf 54}, 1237 (1985);
%R. Bertini {\it et al}, Phys.\ Lett. {\bf B83}, 306 (1979);  R. Bertini
%{\it et al}, Nucl.\ Phys. {\bf A360}, 315 (1981).
%
\bibitem{ioffe1}B.~L. Ioffe,
Nucl.\ Phys.\ {\bf B188}, 317 (1981);
{\bf B191}, 591(E) (1981).
%
\bibitem{ioffe2}B.~L. Ioffe,
Z.\ Phys.\ C {\bf 18}, 67 (1983).
%
\bibitem{reinders1}
L.~J. Reinders, H. Rubinstein, and S. Yazaki,
Phys.\ Rep.\ {\bf 127}, 1 (1985).
%
\bibitem{fetter}A.~L. Fetter and J.~D. Walecka,
{\it Quantum theory of Many-Particle Systems} (McGraw-Hill,
New York, 1971)
%
\bibitem{collins}J.~C.\ Collins, {\it Renormalization} (Cambridge
  Univ.\ Press, New York, 1984).
%
\bibitem{fock1}V.~A. Fock,
Physikalische Zeitschrift der Sowjetunion {\bf 12}, 404 (1937).
%
\bibitem{schwinger1}J. Schwinger,
{\it Particles, Sources, and Fields}
(Addison-Wesley, Reading, 1970), Vol.~I.
%
\bibitem{cronstrom1}C. Cronstr\"om,
Phys.\ Lett.\ {\bf 90B}, 267 (1980).
%
\bibitem{smilga1}A.~V. Smilga,
Yad.\ Fiz.\ {\bf 35}, 473 (1982)
[Sov.\ J.\ Nucl.\ Phys.\ {\bf 35}, 271 (1982)].
%
\bibitem{hubschmid1}W. Hubschmid and S. Mallik,
Nucl.\ Phys.\ {\bf B207}, 29 (1982).
%
\bibitem{novikov3}V.~A. Novikov, M.~A. Shifman,
A.~I. Va\u{\i}nshte\u{\i}n, and V.~I. Zakharov,
Fortschr.\ Phys.\ {\bf 32}, 585 (1984).
%
\bibitem{shifman2}M. Shifman,
Nucl.\ Phys.\ {\bf B173}, 13 (1980).
%
\bibitem{shifman1}M.~A. Shifman, A.~I. Va\u{\i}nshte\u{\i}n,
and V.~I. Zakharov,
Nucl.\ Phys.\ {\bf B147}, 385 (1979);
{\bf B147}, 448 (1979);
{\bf B147}, 519 (1979).
%
\bibitem{ioffe3}B.~L. Ioffe and A.~V. Smilga,
Nucl.\ Phys.\ {\bf B232}, 109 (1984).
%
\bibitem{peshkin}B.~E. Peskin, Phys.\ Lett.\ {\bf 88B}, 128(1979).
%
\bibitem{asymmetry}R. J. Furnstahl, Ohio State University Report
   No.~OSU--0901 (1993).
%
\bibitem{belyaev1}V.~M. Belyaev and B.~L. Ioffe,
Zh.\ Eksp.\ Teor.\ Fiz.\ {\bf 83}, 876 (1982);
{\bf 84}, 1236 (1983)
[Sov.\ Phys.\ JETP {\bf 56}, 493 (1982);
{\bf 57}, 716 (1983)].
%
%
%\bibitem{gluck1}M. Gl\"uck, E. Reya, and A. Vogt,
%Z.\ Phys.\ C {\bf 48}, 471 (1990).
%
%\bibitem{gluck2}M. Gl\"uck, E. Reya, and A. Vogt,
%Z.\ Phys.\ C {\bf 53}, 127 (1992).
%
\bibitem{banerjee}M.~K. Banerjee and J.~B. Cammarata,
Phys.\ Rev.\ D {\bf 16}, 1334(1977).
%
\bibitem{gensini}P.~M. Gensini, Nuovo Cim.\ {\bf 60A}, 221(1980).
%
\bibitem{gasser}J. Gasser, H. Leutwyler, and M.~E. Sainio,
Phys.\ Lett.\ {\bf B253}, 252(1991).
%
\bibitem{leinweber1}D.~B. Leinweber,
Annals of Physics {\bf 198}, 203 (1990).
%%
\bibitem{dorokhov}A.~E. Dorokhov and N.~I. Kochelev,
Z.\ Phys.\ C {\bf 46}, 281 (1990).
%
\bibitem{forkel}H. Forkel and M.~K. Banerjee,
Phys.\ Rev.\ Lett.\ {\bf 71}, 484(1993).
%
\end{references}
\end{document}